\RequirePackage{fix-cm}
\documentclass[twocolumn,epjc3]{svjour3}  
\smartqed  
\RequirePackage{graphicx}
\RequirePackage{lineno}
\usepackage{hyperref}
\usepackage{textgreek}
\usepackage{booktabs}
\usepackage{fixltx2e}
\usepackage{color,soul}

\newcommand{\bbno} {{0\textnu\textbeta\textbeta\xspace}}
\newcommand{\ctsper} {{cts/(keV\(\cdot\)kg\(\cdot\)yr)}}
\journalname{Eur. Phys. J. C}

\usepackage{xspace}
\newcommand{\Gerda} {{\textsc{Gerda}\xspace}}
\newcommand{\LArGe} {{{LArGe}\xspace}}
\newcommand{\I}[2]{$^{#1}$#2}
\newcommand{\tsb}{\textsubscript}
\newcommand{\ts}{\textsuperscript}
\begin{document}

\title{Mitigation of \I{42}{Ar}/\I{42}{K} background for the GERDA Phase II experiment}

\author{A.~Lubashevskiy\thanksref{e1,addr1,addr2}
        \and
        M.~Agostini\thanksref{addr3} 
        \and
        D.~Budj\'{a}\v{s}\thanksref{addr4} 
        \and
        A.~Gangapshev\thanksref{addr1,addr5} 
        \and
        K.~Gusev\thanksref{addr2,addr4,addr6} 
        \and
        M.~Heisel\thanksref{addr1} 
        \and
        A.~Klimenko\thanksref{addr1,addr2} 
        \and
        A.~Lazzaro\thanksref{addr4} 
        \and
        B.~Lehnert\thanksref{addr7,addr10} 
        \and
        K.~Pelczar\thanksref{addr8} 
        \and
        S.~Sch\"{o}nert\thanksref{addr4} 
        \and
        A.~Smolnikov\thanksref{addr1,addr2} 
        \and
        M.~Walter\thanksref{addr9} 
        \and
        G.~Zuzel\thanksref{addr8} 
}

\thankstext{e1}{e-mail: lav@nusun.jinr.ru}


\institute{Max Planck Institut f{\"u}r Kernphysik, Heidelberg, Germany \label{addr1}
           \and
           Joint Institute for Nuclear Research, Dubna, Russia\label{addr2}
           \and
            Gran Sasso Science Institute, L'Aquila, Italy\label{addr3}
           \and
           Physik Department E15, Technische  Universit{\"a}t M{\"u}nchen, Munich, Germany\label{addr4}
           \and
           Institute for Nuclear Research of the Russian Academy of Sciences, Moscow, Russia\label{addr5}
             \and
           Russian Research Center Kurchatov Institute, Moscow, Russia\label{addr6}
           \and
           Institut f{\"u}r Kern- und Teilchenphysik Technische Universit{\"a}t Dresden, Dresden, Germany\label{addr7}
           \and
           Institute of Physics, Jagellonian University, Cracow, Poland\label{addr8}
           \and
           Physik Institut der Universit{\"a}t Z{\"u}rich, Z{\"u}rich, Switzerland\label{addr9}
           \and
           \textit{Present address:} Physics Department, Carleton University, Ottawa, Canada\label{addr10}
}

\date{Received: date / Accepted: date}

\maketitle

\begin{abstract}

Background coming from the \I{42}{Ar} decay chain is considered to be one of the most relevant for the \Gerda{} experiment, which aims to search of the neutrinoless double beta decay of \I{76}{Ge}. The sensitivity strongly relies on the absence of background around the Q-value of the decay. Background coming from \I{42}{K}, a progeny of \I{42}{Ar}, can contribute to that background via electrons from the continuous spectrum with an endpoint of 3.5 MeV. Research and development on the suppression methods targeting this source of background were performed at the low-background test facility \LArGe{}. It was demonstrated that by reducing \I{42}{K} ion collection on the surfaces of the broad energy germanium detectors in combination with pulse shape discrimination techniques and an argon scintillation veto, it is possible to suppress the \I{42}{K} background by three orders of magnitude. This is sufficient for Phase~II of the \Gerda{} experiment.
\\\keywords{\I{42}{K} background \and \I{42}Ar \and nylon mini-shroud \and liquid argon veto \and \LArGe \and BEGe}
\end{abstract}

\section{Introduction}

In \Gerda{} bare germanium diodes immersed in liquid argon (LAr) are used both as the source and the detector of the neutrinoless double beta (\bbno{}) decay of \I{76}{Ge}~\cite{GERDA}. Broad Energy Germanium (BEGe) detectors~\cite{Canberra} were introduced to \Gerda{} in Phase~I and dominate the Phase~II detector inventory. BEGe detectors in \Gerda{} have better energy resolution than semi-coaxial detectors and a significantly thinner dead layer~\cite{GERDAbkg}. The latter makes them prone to surface events, from the beta particles, which can penetrate the dead layer. This kind of background is considered to be one of the most dangerous background sources in \Gerda{}~Phase~II.  

\begin{figure}
  \begin{center}
    \includegraphics[width=0.75\linewidth]{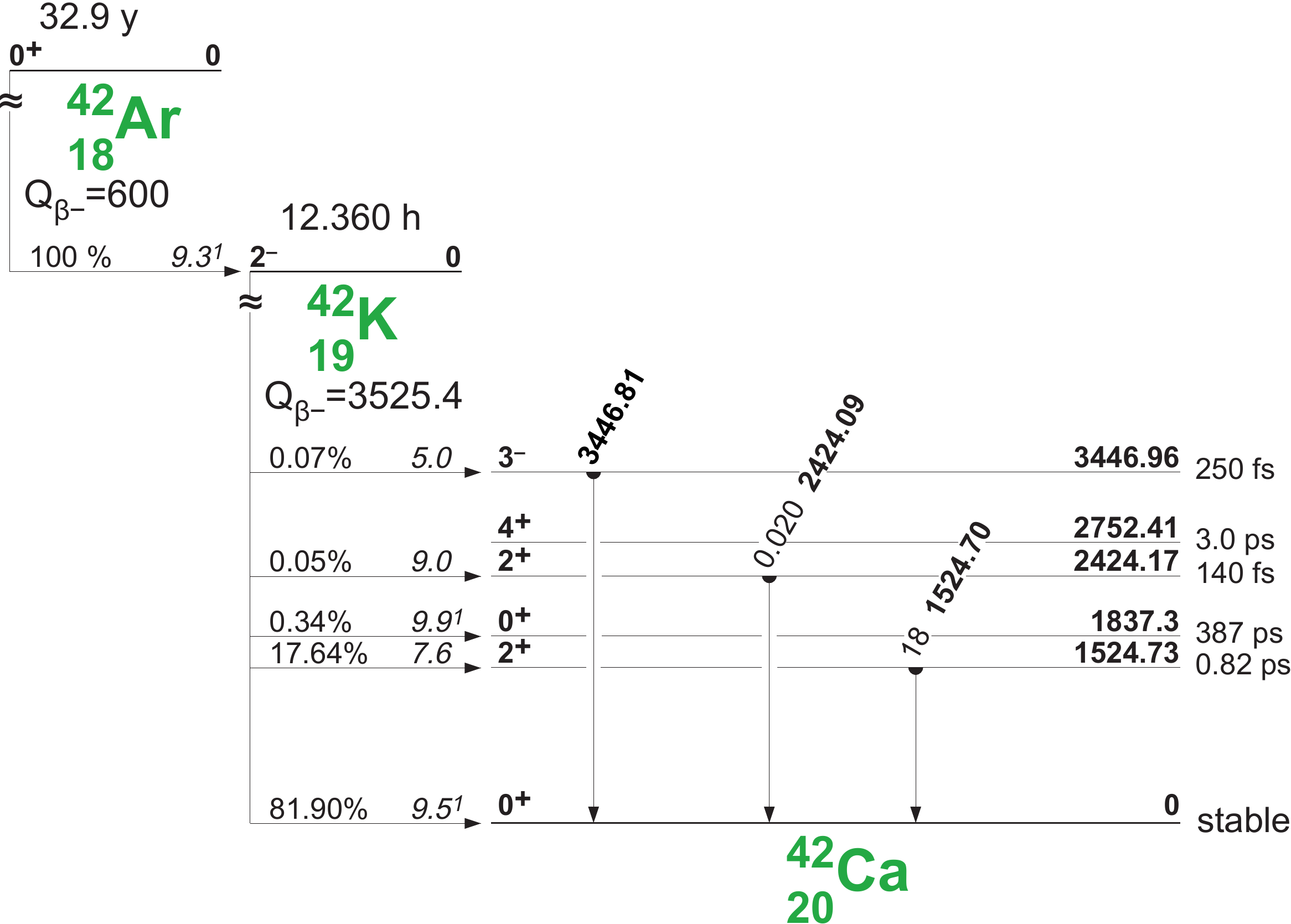}
    \caption{\label{fig:decay}Simplified decay scheme of \I{42}{Ar}~\cite{ToI}. The 1525\,keV gamma line is used to identify \I{42}{K}.}
  \end{center}
\end{figure}

LAr serves as a detector coolant and as passive shielding against external radiation. Background events can be additionally suppressed by a LAr veto, detecting scintillation light generated in the liquid argon. Cosmogenic \I{42}{Ar}, present in natural argon, decays into \I{42}{K}. \I{42}{K} is a beta emitter (Q\tsb{\textbeta} = 3525\,keV) with a half-life of 12.3\,h (see Fig.~\ref{fig:decay}).
\label{sec:intro}
The region of interest (ROI) of the \bbno{} decay is centered around Q\tsb{\textbeta\textbeta} = 2039\,keV~\cite{qbb}. 
High energy electrons E~$\ge$~Q\tsb{\textbeta\textbeta} from \I{42}{K} beta decays occurring very close to the detector surface, can increase the background level in the ROI.
 
During the first commissioning runs of \Gerda{}~Phase~I it was found that the intensity of the 1525\,keV gamma line from \I{42}{K} is much higher than expected from homogeneous distribution of \I{42}{Ar} assuming a natural abundance of $\textless$43\,\textmu Bq/kg~\cite{Ash03}. A similar conclusion emerged from the measurements in the \LArGe{} test facility~\cite{LArGe}, operated in the framework of \Gerda. The measurements~\cite{GERDAbkg} show that the enhancement of \I{42}{K} background can be attributed to an accumulation effect: after the decay of their mother isotope, \I{42}{K} ions retain their positive charge long enough to move in the electric field of the germanium detectors~\cite{Pelczar2016}. Consequently, the \I{42}{K} distribution in liquid argon is not homogeneous.

Enclosing a detector string with a cylinder made from thin copper foil (called mini-shroud, MS) it is possible to considerably suppress the background contribution coming from the \I{42}{Ar} decay chain~\cite{GERDA}. The mini-shroud screens the electric field of the germanium detectors and creates a mechanical barrier preventing the collection of the \I{42}{K} ions on detector surfaces and thus decreasing the \I{42}{K} background level. However, the copper MS could not be used in Phase~II: the main reason is that argon scintillation light generated inside the copper MS would be blocked from detection by the LAr veto, thus the veto efficiency would be seriously impaired. Another reason is that the radiopurity of the copper MS used in \Gerda{} Phase~I does not suffice the more stringent radiopurity requirements of Phase~II. Therefore, it was decided to develop a new mini-shroud. Results from this development and detailed tests of its \I{42}{Ar} mitigation capability are described here. BEGe detectors have powerful pulse shape discrimination (PSD) capability which allows to mitigate surface events from \I{42}{K}. The suppression capability of \I{42}{K} background events by the PSD has been tested in the current investigation. The suppression factors (SF) obtained with passive and active methods are presented.

\section{Experimental}
\label{sec:exper}
%
\subsection{Development of a transparent mini-shroud}
\label{sec:devMS}
%
 \I{42}{K} background can be suppressed e.g. by modifying the electric field produced by the high voltage applied to detectors, or by a mechanical barrier. Several applications were tested for \Gerda{}(mini-shroud made of copper mesh, copper plate, plastic scintillator, etc.) and proved to work well in liquid argon. The solution described here is the mini-shroud made from a transparent plastic foil. The MS does not screen the E-field of the detector, but as a mechanical barrier for \I{42}{K} ions prevents their drift towards the germanium detector. The mini-shroud encapsulates the detector, but is not tight to prevent liquid argon from pouring inside during the immersion. The collected \I{42}{K} atoms decay on the foil's surface at a distance of several millimeters from the detector, thus beta particles are attenuated by liquid argon.

A thin nylon film (thickness 125\,\textmu m) was chosen as the construction material. It is robust, durable and flexible. It has good transparency for the visible light and exhibits a very low intrinsic radioactivity. Several pieces of nylon film were provided by Princeton University. These are left-overs from the material used in Borexino to construct the Inner Vessel~\cite{bor_vessel}. The radioactive impurities are only 2\,ppt for \I{238}{U} and 4\,ppt for \I{232}{Th}~\cite{bor_catt}. 

Scintillation light emitted by liquid argon (128\,nm) must be converted to wavelengths suitable for scintillation light detectors (sensitive range is about 250 -- 500\,nm). Thus the nylon mini-shroud (NMS) was covered with a wavelength shifter (WLS). It allows light to pass through the nylon film as it is opaque for deep ultraviolet radiation (below 300\,nm). The performance of the WLS coating, its robustness and stability in time were also investigated in various tests. These studies include an optimization of the composition and thickness of the WLS coating.

The performance of the WLS coating was tested with a UV spectrophotometer (Cary Eclipse) at the Max-Planck-Institut f\"{u}r Kernphysik (MPIK) in Heidelberg. Direct investigation of the shifting of argon scintillation light at 128\,nm was not possible since no light source in this far UV region was available. Instead, a qualitative comparison was made with the emission light of the spectrophotometer at 200\,nm.
  
\begin{figure}
  \begin{center}
    \includegraphics[width=0.9\linewidth]{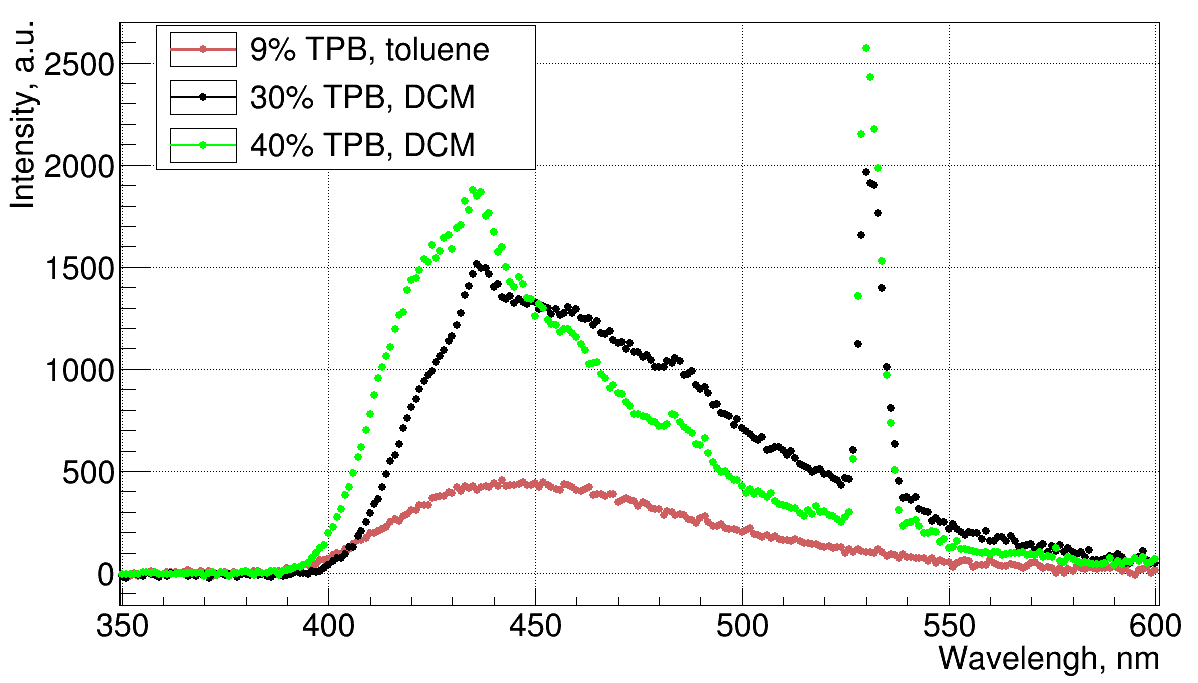}
    \caption{\label{fig:aging1}Fluorescence spectra of the coated nylon samples with different composition of WLS  measured with a Cary Eclipse UV spectrophotometer. The peak at 530 nm is due to scattered light from the excitation beam.}
  \end{center}
\end{figure}

The initial WLS solution was prepared by dissolving tetra-phenyl-buthadiene (TPB) with polystyrene (PS) in a ratio of 1:10 in toluene (recipe previously used for coating of the VM2000 reflection foil in \LArGe{}~\cite{Peiffer_PhD}). To improve the performance of the nylon mini-shroud we increased the concentration of TPB and replaced toluene by dichlormethane (DCM) as the latter is used for the production of the reflector foils in \Gerda{} Phase~II~\cite{GERDAfoils}. A higher concentration of TPB in the polystyrene matrix increases the intensity of emitted light (see Fig.~\ref{fig:aging1}).
However, too high TPB concentration decreases the mechanical stability of the coating. Finally, a TPB concentration in the range of 30\% - 40\% in the polystyrene matrix proved to be a good balance between mechanical stability and light emission efficiency. The coating was applied by brushing both sides of nylon samples. The amount of deposited WLS (typically 0.3\,mg per 1\,cm\ts{2}) was estimated by weighting the nylon before and after coating. 
 
\begin{figure}
  \begin{center}
    \includegraphics[width=0.9\linewidth]{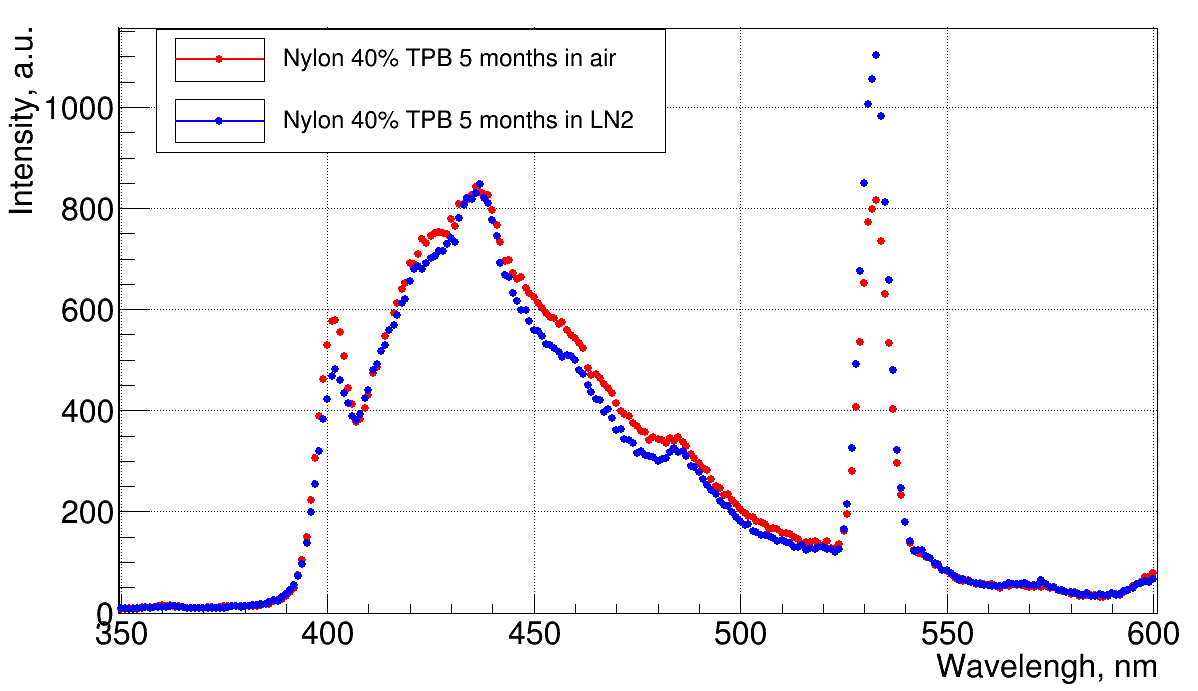}
    \caption{\label{fig:aging2}Fluorescence spectra of the coated nylon samples
      stored in liquid nitrogen for 5 months (blue) and in air (red).}
  \end{center}
\end{figure}

Various tests were performed in order to understand the usability of such a foil in \Gerda{}. Immersion of the nylon film in liquid nitrogen showed that it keeps flexibility at low temperatures and does not become brittle even after several weeks spent in cryogenic liquid. Long term stability of the coated nylon foils was checked in liquid nitrogen (five months) and liquid argon (two months in the \LArGe{} cryostat). No visible deterioration of the coated surfaces was found. Changes in optical properties of the WLS were checked with a UV spectrophotometer. Fig.~\ref{fig:aging2} shows a comparison of samples kept for five months in liquid nitrogen (blue dots) and in air (red dots). 
No significant difference between the two fluorescence spectra was observed.

\begin{figure}
  \begin{center}
    \begin{minipage}{0.49\linewidth}
      \center{\includegraphics[width=0.9\linewidth]{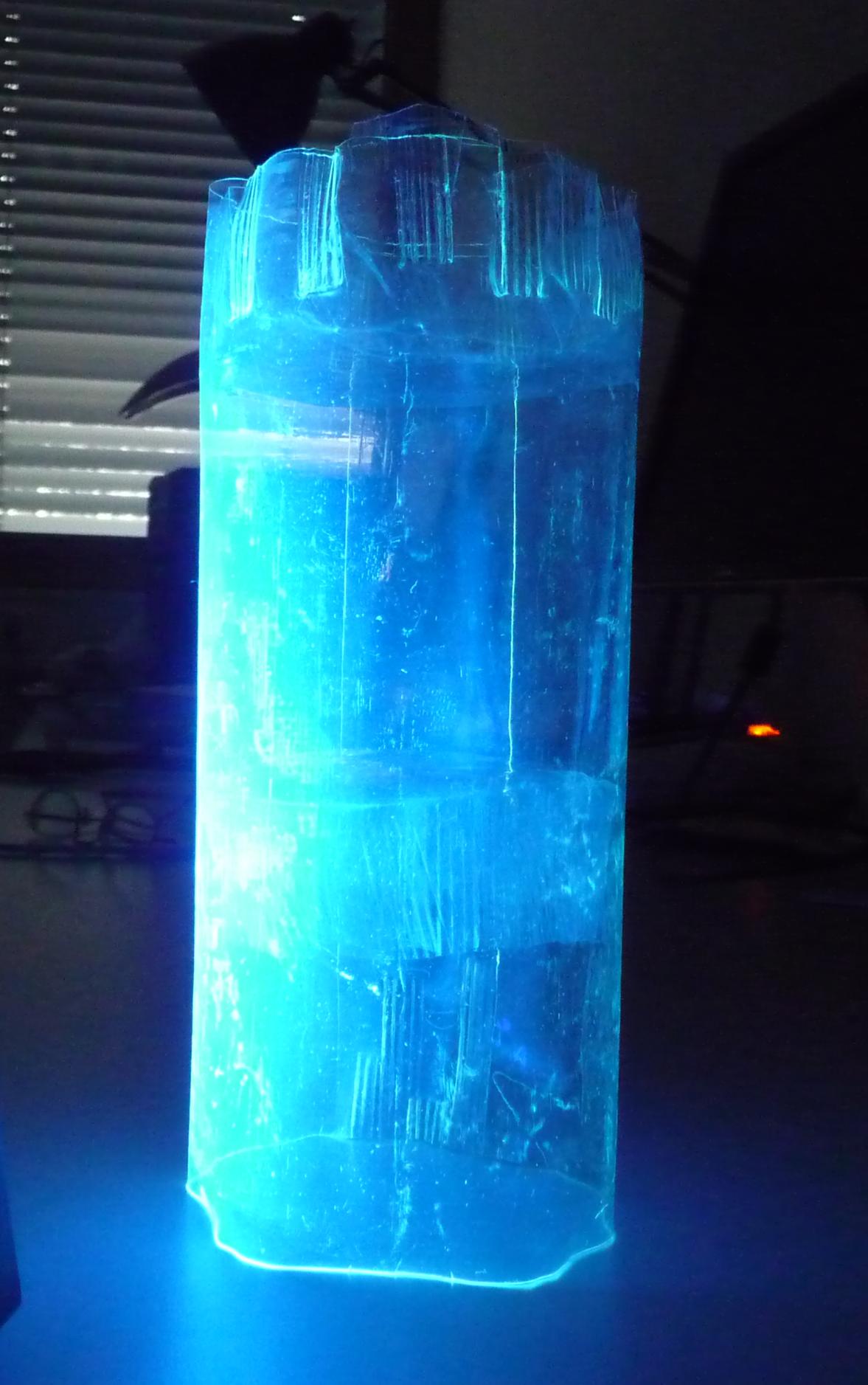}}
    \end{minipage}
    \hfill
    \begin{minipage}{0.49\linewidth}
      \center{\includegraphics[width=0.9\linewidth]{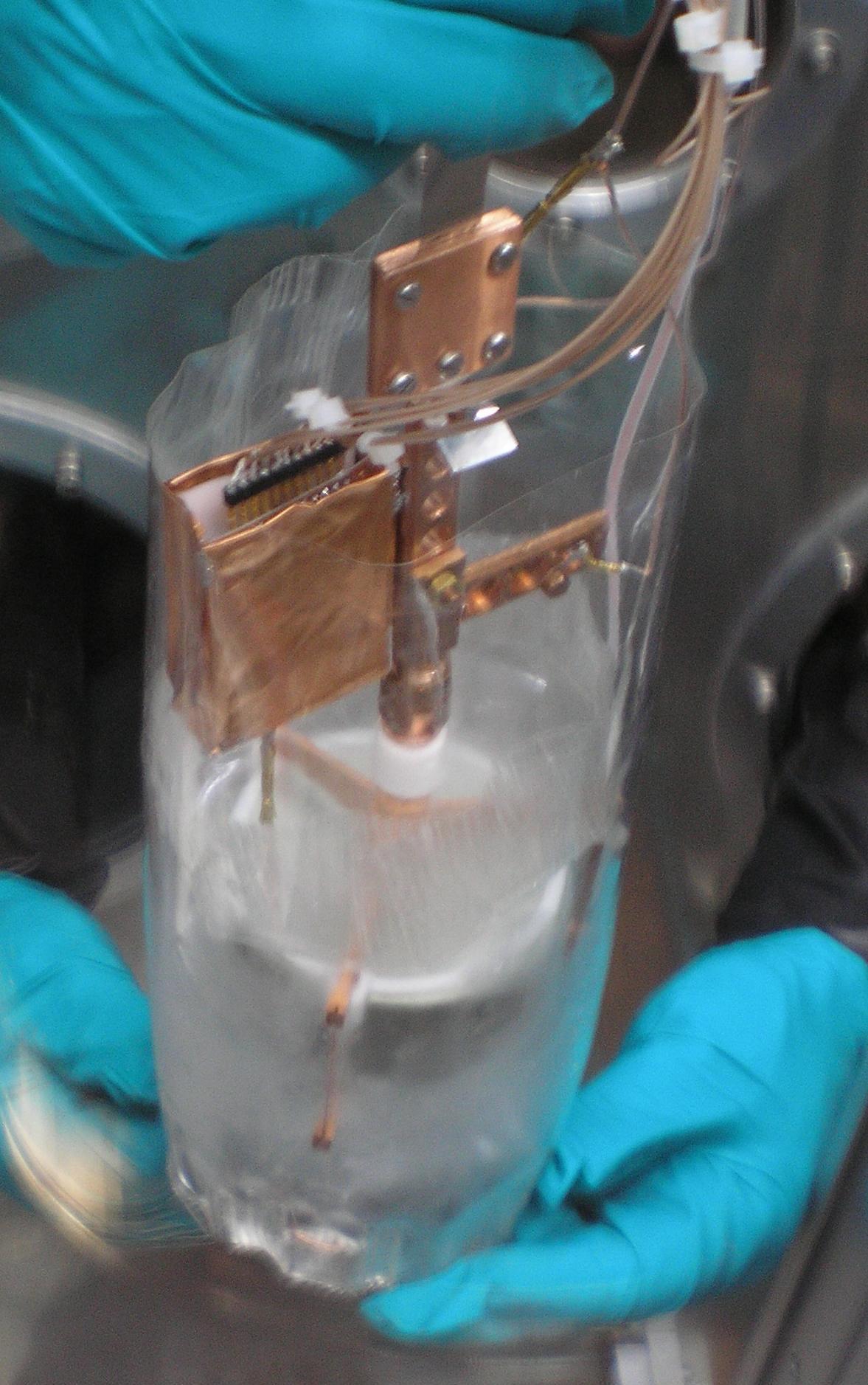}}
    \end{minipage}
    \caption{\label{fig:nylon1}Left: photo of the first version of a nylon mini-shroud coated with WLS (the photos were taken in UV light). The nylon sheets were connected by welding. Right: NMS in assembly with a bare BEGe detector inside the lock of \LArGe.}
  \end{center}
\end{figure}

To construct the NMS fully covering the detector (or a string of detectors) it is necessary to connect several nylon sheets together to form a cylindrical shape. Such connections should be radioactively clean and robust enough to survive manual handling and immersion into liquid argon. The first version of the mini-shroud was created by welding (see Fig.~\ref{fig:nylon1}) not requiring any additional material, but the connections were brittle and easily damaged during handling.
Finally, nylon pieces were glued using a solution of resorcinol in ethanol and water with sodium meta-bisulphite~\cite{borexino_glue}. It was confirmed that the connections are very robust and suitable for cryogenic operations. The glued NMS was placed around a dummy detector string (aluminum dummies of detectors and holders) and it was submerged in liquid nitrogen, showing that it is robust enough and can keep its shape during the slow immersion process ($\sim$1 hour) of the detector strings in \Gerda{}. To minimize the risk of contamination, all preparations were performed in the clean room. Before the insertion into liquid argon, the NMS went through several outgassing cycles. 

\subsection{Contribution of the NMS to the \Gerda{} background index}
\label{sec:devMS}

The content of radio-impurities of coated nylon films was checked by ICP-MS to verify that no significant background
contribution is added to GERDA~\cite{icp_ms}. Results of the measurements are summarized in Tab.~\ref{tab:icpms}. 
\begin{table}
  \begin{center}
    \caption{\label{tab:icpms}Radiopurity of the components of the NMS from ICP-MS measurements.}
\begin{tabular}{lrrr}
  \toprule
Components & Mass per & U & Th \\[2mm]
& NMS (g) & (ppt) & (ppt) \\[2mm]
  \hline \\ [-2.0ex]
TPB & $\sim$0.15 & 10 & 9 \\
Polystyrene & $\sim$0.34 & \textless5 & 10  \\
Glue &  - &\textless10 & \textless10   \\
Uncoated nylon & 27.6 &\textless10 & \textless15   \\
\hline \\ [-2.5ex]
Coated nylon  & 21.5 & 11 & 18\\
Glued nylon & 6.5 & 38 & 39  \\
\hline\\ [-2.5ex]
NMS & 28.1 & 6.1 \textmu Bq & 2.6 \textmu Bq\\
   \bottomrule
\end{tabular}
\end{center}
\end{table}
The difference in the radioactive contamination of the starting foil material and the aggregated contamination of the final NMS may have been introduced during the preparation of the samples. The nylon cylinder for each \Gerda{}~Phase~II detector string typically has a height of 430 mm and a diameter of 103 mm. The mass of the NMS is about 28\,g. We expect about 6.1\,\textmu Bq of \I{238}{U} and about 2.6\,\textmu Bq of \I{232}{Th} reside in a single NMS. 

\begin{figure}
  \begin{minipage}{1\linewidth}
    \center{\includegraphics[width=0.9\linewidth]{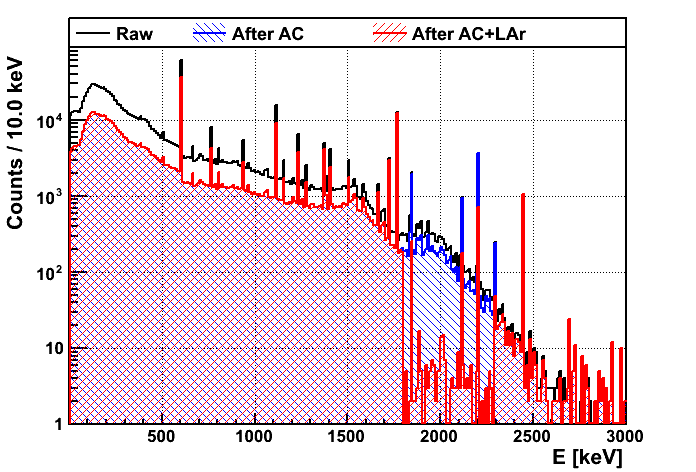}}
  \end{minipage}
  \hfill
  \begin{minipage}{1\linewidth}
    \center{\includegraphics[width=0.9\linewidth]{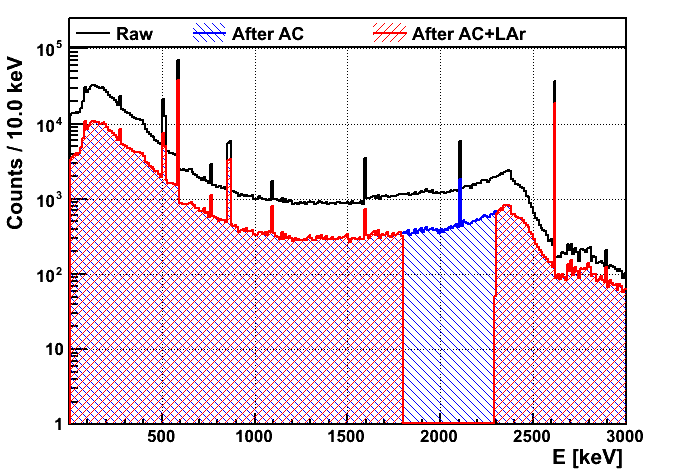}}
  \end{minipage}
  \caption{\label{fig:sim_spectrum}Simulated energy spectra of germanium detector taken with \I{214}{Bi} (top) and \I{208}{Tl} (bottom) sources. The simulated spectra before cuts are shown in black, after the detector anti-coincidence (AC) in blue and after applying the LAr scintillation veto cut in red~\cite{phd_bjoern}. Simulation of the anti-coincidence spectrum was only performed near Q\tsb{\textbeta\textbeta}, to save computation time. }
\end{figure}

To prove that this radiopurity level is acceptable, a detailed simulation of the NMS background for the \Gerda{} Phase II setup was performed. This Monte Carlo simulation is based on the MaGe framework~\cite{MAGE} with the preliminary \Gerda{} Phase II geometry, using 7 detectors strings with one NMS each, under the assumption of equilibrium in the U/Th chains~\cite{phd_bjoern}. The simulation code is validated with the \Gerda{} calibration data. Radioactive isotopes were homogeneously distributed inside the NMS material. The simulated energy spectra of germanium detectors taken with \I{214}{Bi} and \I{208}{Tl} sources are shown in Fig.~\ref{fig:sim_spectrum}.
Tracking of optical photons and the evaluation of the LAr veto suppression were only performed in the energy region of 1800-2300 keV, to save computation time. Suppression factors of 33$\pm$4 and $\ge$ 2000 are obtained for \I{214}{Bi} and \I{208}{Tl} respectively. The large suppression is due to the thinness of the NMS which allows the betas from \I{208}{Tl} and \I{214}{Bi} to escape the nylon and deposits a large fraction of their energy in the LAr. This creates a large scintillation signal which can be easily vetoed. The estimated contributions to the background index (BI) for \I{214}{Bi} are 1.8$\times$10$^{-4}$\,\ctsper{} before and 5.3$\times$10$^{-6}$\,\ctsper{} after applying the LAr veto. For \I{208}{Tl} these values are 2.7$\times$10$^{-4}$\,\ctsper{} and $\textless$1.4$\times$10$^{-7}$\,\ctsper{} before and after the LAr veto, respectively. The combination of these values gives the total contribution of the NMS to the BI in \Gerda{} Phase II of about 4.6$\times$10$^{-4}$\, \ctsper{} and 5.3$\times$10$^{-6}$\, \ctsper{} before and after the LAr veto, respectively. These values are well below the \Gerda{}~Phase~II requirements (total BI\textless 1$\times$10$^{-3}$\, \ctsper{}).
%
\subsection{Experimental setup}
\label{sec:expersetup}
%
Detailed investigations of the \I{42}{K} background were performed in the \LArGe{} test facility~\cite{LArGe}. The setup allows to operate bare germanium detectors submerged in 1\,m$^3$ of liquid argon. The cryostat is equipped with nine photomultiplier tubes (PMT) in order to suppress background events by detecting scintillation light in liquid argon, occurring in coincidence with germanium detector events (LAr veto). The \LArGe{} setup is located at the Laboratori Nazionali del Gran Sasso (LNGS) Underground Laboratory at 3800\,m w.e. depth. An achieved background index in \LArGe{}~\cite{LArGe} is close to to the level of \Gerda{}~Phase~I. Hence, an investigation of low background sources is possible.

\begin{figure}
  \begin{center}
    \includegraphics[width=0.9\linewidth]{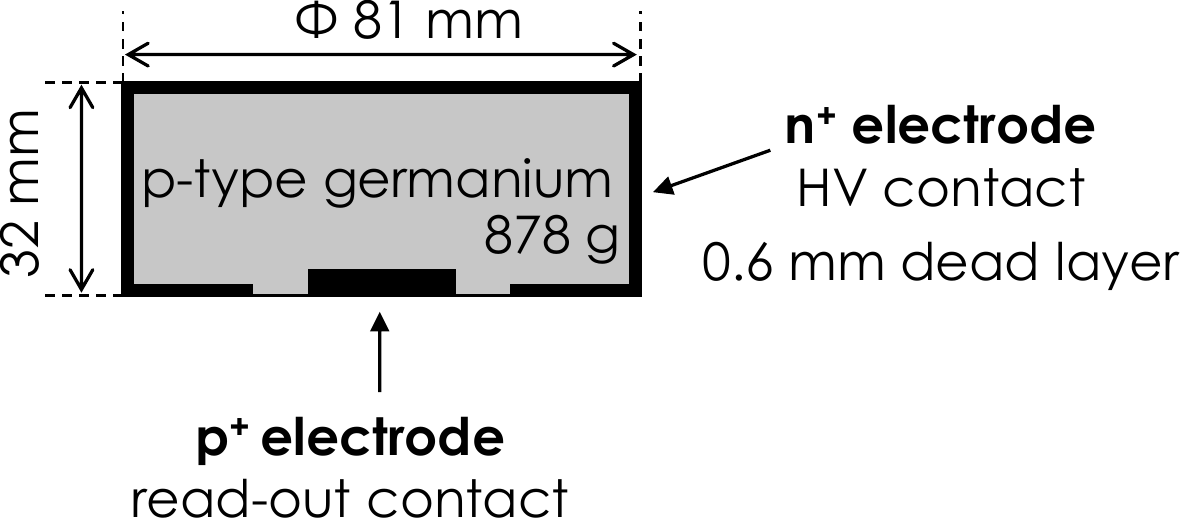}
    \caption{\label{fig:bareBEGe1}Schematic drawing of the bare BEGe detector used in the study~\cite{psd}.}
  \end{center}
\end{figure}

The measurements in \LArGe{} were performed with a BEGe detector produced by Canberra~\cite{canberra}. The scheme of the detector and its dimensions are shown in Fig.~\ref{fig:bareBEGe1}. 
It is a p-type diode with the mass of 878\,g. The thickness of the dead layer is estimated to be 0.6\,mm. The detector mounting procedure and the holder design is the same as in \Gerda{} Phase~I~\cite{GERDA}, i.e. usage of low-mass and radio-pure materials close to the detector is implied. A \Gerda{} Phase I custom made charge-sensitive preamplifier (CC2) with integrated J-FET transistor and RC feedback components is used for the signal conditioning~\cite{cc2}. Signals are digitized with a Flash-ADC (Struck SIS3301 VME, 14 bit, 100 MHz) card.

In order to enhance the \I{42}{K} signal, two \I{42}{Ar} sources with activities of (5.18 $\pm$ 0.91)\,Bq and (79 $\pm$ 15)\,Bq were produced by irradiating a cell filled with gaseous \I{nat}{Ar} at the accelerator operated in the Maier-Leibnitz Laboratory (MLL) of the Technische Universit\"{a}t M\"{u}nchen (TUM). The sources were introduced into the \LArGe{} cryostat and homogeniously distributed inside LAr. With the enhanced signal rate from \I{42}{K} it is possible to investigate the behavior of \I{42}{K} ions and to test different background suppression methods with better counting statistics. The long measurement without NMS and measurement with a first version of the NMS were performed after the insertion of the source with lower activity. Measurements with the second, final version of the NMS were done with both sources inside LAr. The \I{42}{Ar} activity of these sources increased the count rate of the 1525\,keV gamma peak (with respect to the measurements with natural argon) by factors of about 40 and 200, respectively. The increase is not exactly proportional to the activity due to the evaporation of liquid argon during refilling of the cryostat. 
%

\section{Measurements of suppression factors}
\label{sec:res}
The energy region of 1540 -- 3000\,keV ("beta region") contains predominantly surface events from beta particles (see the \I{42}{K} decay scheme in Fig.~\ref{fig:decay}) and overlaps the Q\tsb{\textbeta\textbeta}$\pm$200\,keV energy region. The energy region 1520--1530 keV mostly contains events from the 1525 keV gamma line. 

The suppression factor in a certain energy region expresses the ratio of events in the unsuppressed ($N_{0}$) versus the suppressed ($N_{S}$) spectrum. For measurements with LAr veto in order to be independent of the source strength the LAr acceptance ($\varepsilon_{LAr}$) is taken into account, hense SF = $ (N_{0}-B_{0})\cdot \varepsilon_{LAr}/(N_{S}-B_{S})$, $B_{0}$ and $B_{S}$ are the background levels before and after suppression respectively. 
 
%
\subsection{Suppression of \I{42}{K} background by NMS}
\label{sec:suppr42K}

\begin{figure}
  \begin{center}
    \begin{minipage}{0.49\linewidth}
      \center{\includegraphics[width=1\linewidth]{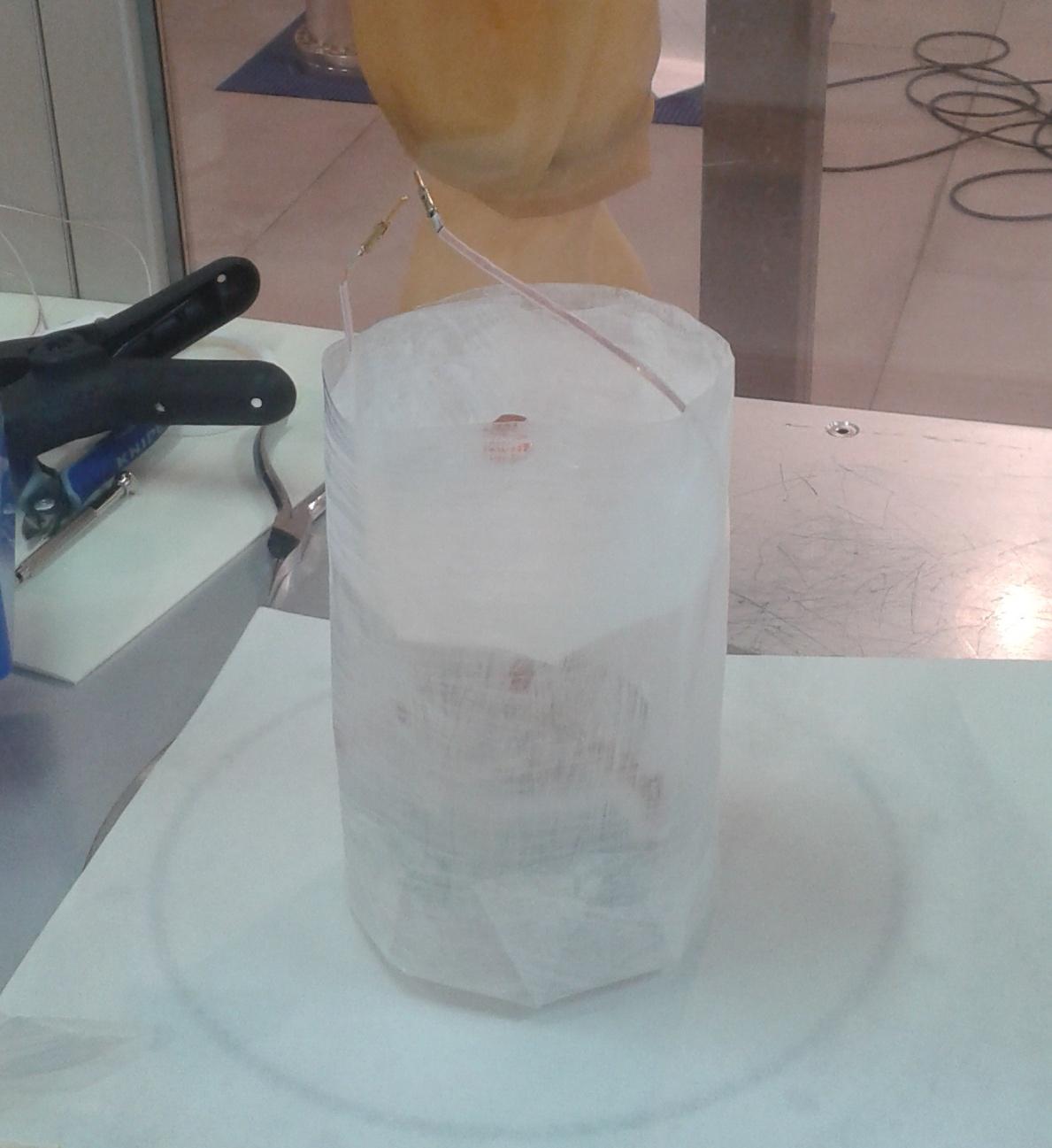}}
    \end{minipage}
    \hfill
    \begin{minipage}{0.49\linewidth}
      \center{\includegraphics[width=1\linewidth]{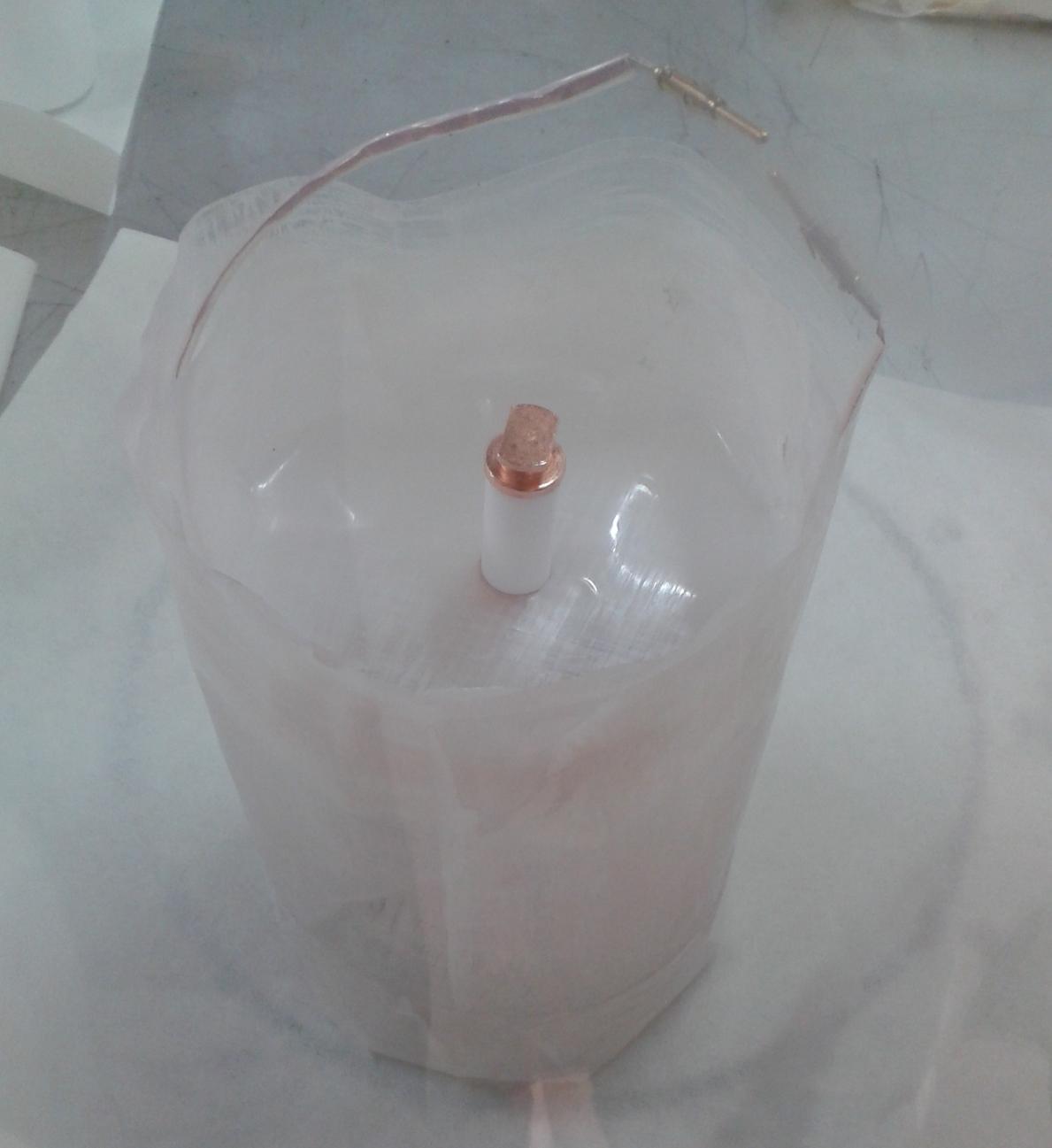}}
    \end{minipage}
    \caption{\label{fig:nylon_glued} Glued version of the NMS around the bare BEGe detector inside the glove box prior incertion into \LArGe{}.}
  \end{center}
\end{figure}

\begin{figure}
\begin{center}
\includegraphics[width=1\linewidth]{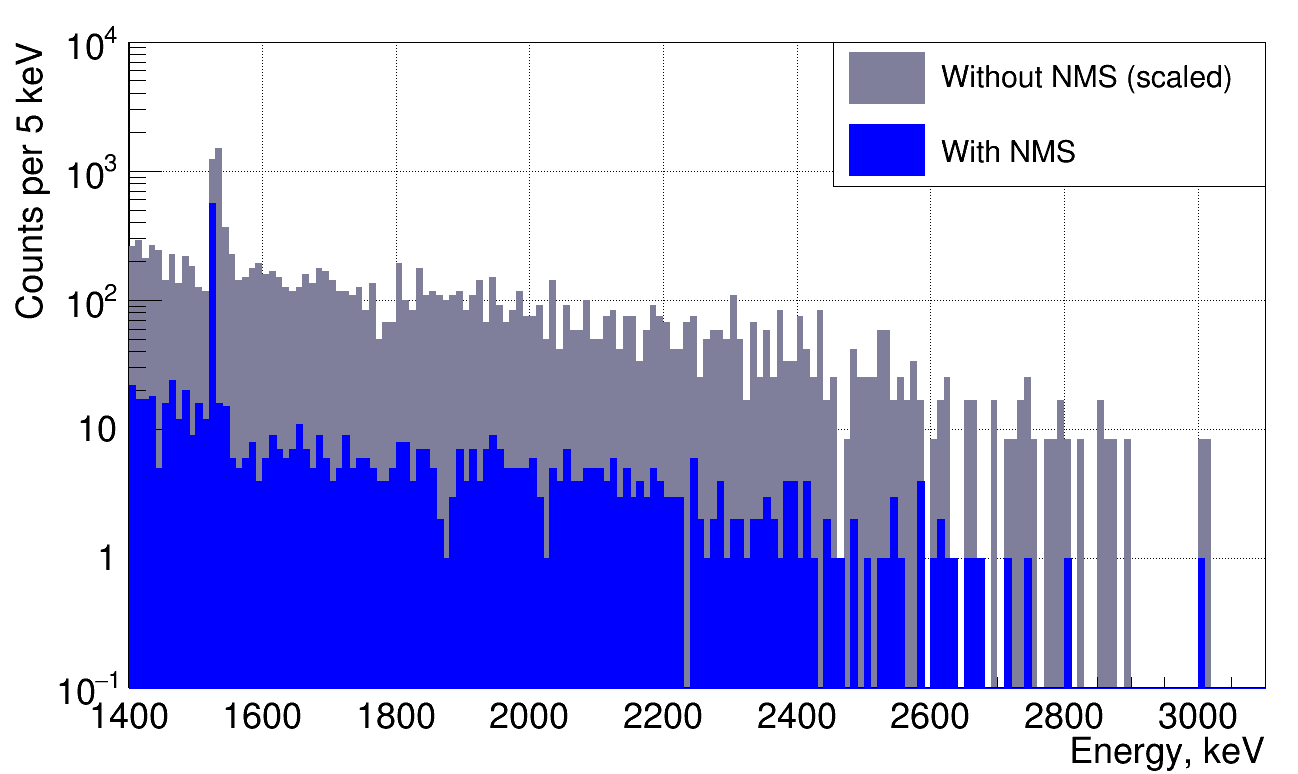}
  \caption{\small\label{fig:spe_nms_bege} Normalized energy spectra of the bare BEGe detector without NMS (grey area) and with NMS (blue).  
}
\end{center}
\end{figure}

The BEGe detector inside the NMS (Fig.~\ref{fig:nylon_glued}) was submerged into the \LArGe{} cryostat. 
No deterioration of the detector's performance was found when operated with the NMS. A typical energy resolution of the detector is about 2.9\,keV at the 2.6\,MeV of the \I{208}{Tl} gamma line, similar to measurements without NMS. The energy spectra with and without NMS are shown in Figure~\ref{fig:spe_nms_bege}. 

\begin{figure}
\begin{center}
\includegraphics[width=0.9\linewidth]{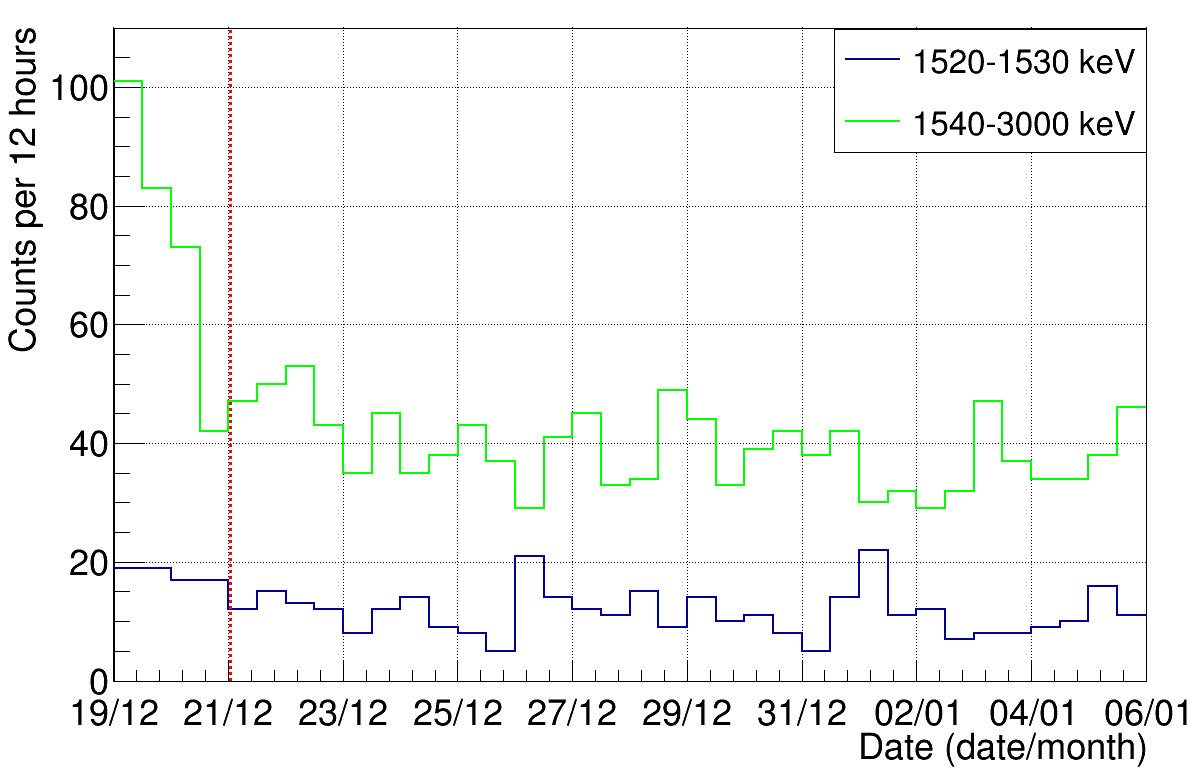}
  \caption{\small\label{fig:time_both_bare}
Number of counts in the energy region 1520-1530~keV (blue) and 1540-3000~keV (green) in time for measurements of the detector without NMS in LArGe. The vertical red line indicates a period of the first two days which is not taken into account in the calculations. 
}
\end{center}
\end{figure}

The measurement without NMS was taken for a shorter time, therefore its spectrum is scaled to the NMS measurement time. Tab.~\ref{tab:nms} shows results from the calculation of the SF in the "beta region" using the ratio of count rates with and without NMS.
\begin{table}
\begin{center}
\caption{\small\label{tab:nms}
Count rates in the different energy regions with and without NMS. SF in the "beta region" is calculated taking into account long-term decrease of the count rates. 
}
\begin{tabular}{crrc}
\toprule
     & 1520-1530 keV  & 1540-3000 keV\\
    &  [cts/(kg d)] & [cts/(kg d)]  \\  
\hline
 without NMS & 99(10) & 332(18)   \\
with NMS    & 29.0(2.3) & 17.1(1.8) \\
\hline
SF  &   & 14.3(2.1) \\
\bottomrule
\end{tabular}
\end{center}
\end{table}
The first two days of data taking are not taken into account in order to reduce the influence of previously accumulated \I{42}{K}. An additional decrease of the \I{42}{K} count rate after first two days (see Figure~\ref{fig:time_both_bare}) is also taken into account during calculation of the SF (henceforth referenced as "long-term correction").
Thus the SF is estimated to be 14.3(2.1). 
%
\subsection{Suppression of \I{42}{K} background by PSD and LAr veto}
\label{sec:suppr42K}

A strong weighting potential close to the read-out electrode in the BEGe-type detectors enables a powerful pulse shape discrimination (PSD) of the acquired signals~\cite{psd_sim}. The PSD used in this work is based on the method described in~\cite{psd}. It was developed for the discrimination between single-site events (SSE) and multi-site events (MSE) using differences in the amplitudes of current pulses with the same energy. The \bbno{} events are expected to be SSE due to a rather short path length of beta particles in the germanium detectors. In contrast, photons will often undergo several scatterings and are mostly MSE. Surface events exhibit a slower rising edge of the pulse due to diffusion of the electric charge into the n\ts{+} surface areas (n\ts{+} contact surface pulses, NSP). In consequence, these events have smaller amplitude to energy (A/E) ratio compared to events in the bulk of the detector. The events that deposit energy near the p\ts{+} contact of the detector (p\ts{+} contact pulses, PCP) have higher amplitudes than SSE, and they can be rejected as well.
\begin{figure}
  \begin{center}
    \includegraphics[width=1\linewidth]{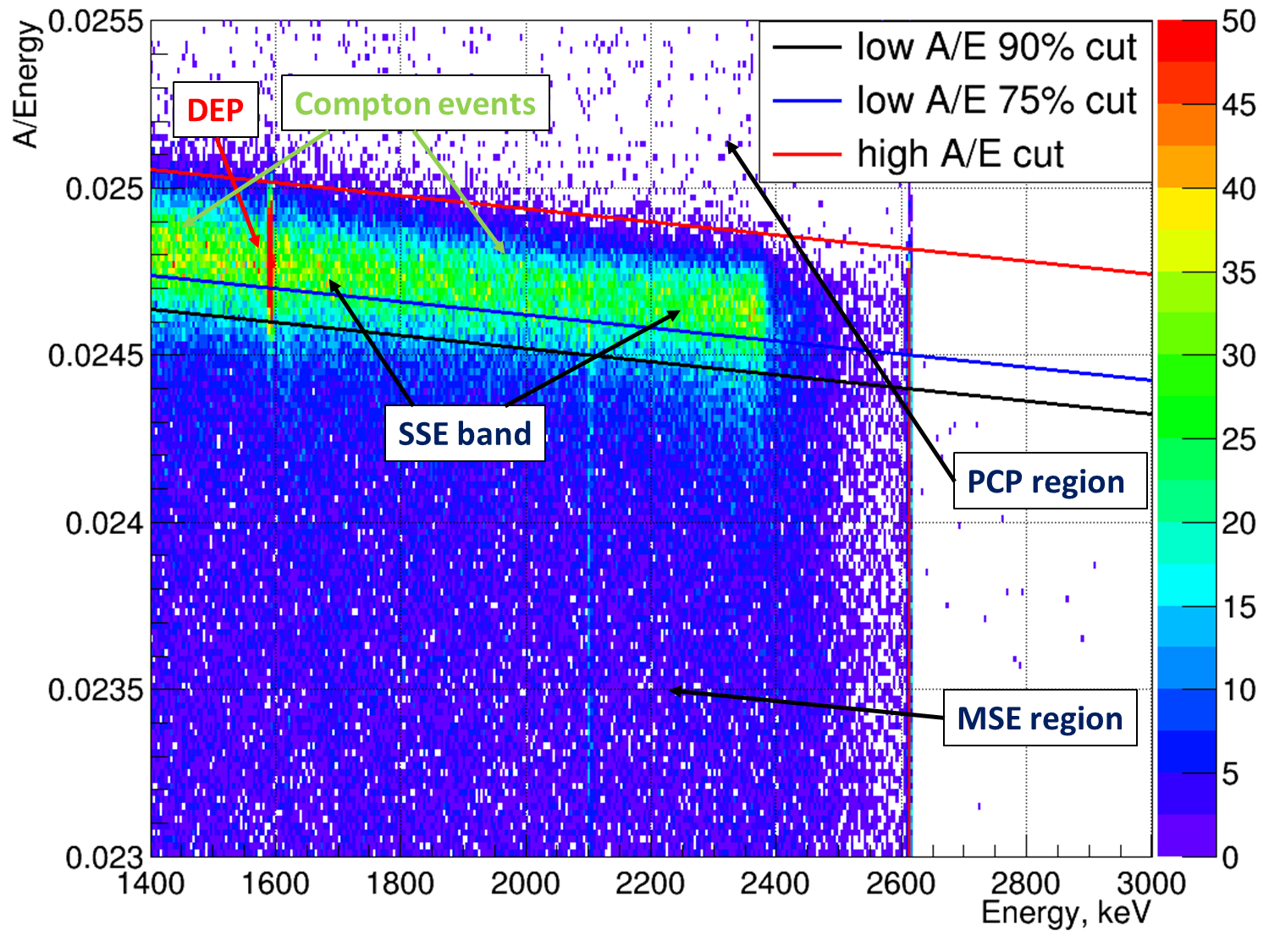}
    \caption{\label{fig:ae_th228} Density diagram of the A/E parameter as a function of energy obtained in the measurements with the BEGe detector in \LArGe{} with \I{228}{Th} source. Lines represent parameters of the PSD cuts obtained from this spectrum using events in the DEP and the Compton continuum.}
  \end{center}
\end{figure}

The PSD cut parameters were determined from calibration data taken with a \I{228}{Th} source. Events in the double escape peak (DEP) of the 2.6\,MeV gamma line of \I{208}{Tl} are predominantly single-site with a shape similar to \bbno{} events. The low cut is usually determined in order to keep 90\% of the events in the DEP. Fig.~\ref{fig:ae_th228} demonstrates such PSD cut obtained for the measurement with the BEGe detector in \LArGe{} (rejected events are in the MSE region below the black line). The slope of the lines is determined from the distribution of the Compton scattering events. The p+ events are rejected by applying a high cut (these events are above the red line in Fig.~\ref{fig:ae_th228}). After applying low and high cuts 88\% of the events are kept in the DEP. 

\begin{figure}
\begin{center}
    \includegraphics[width=1\linewidth]{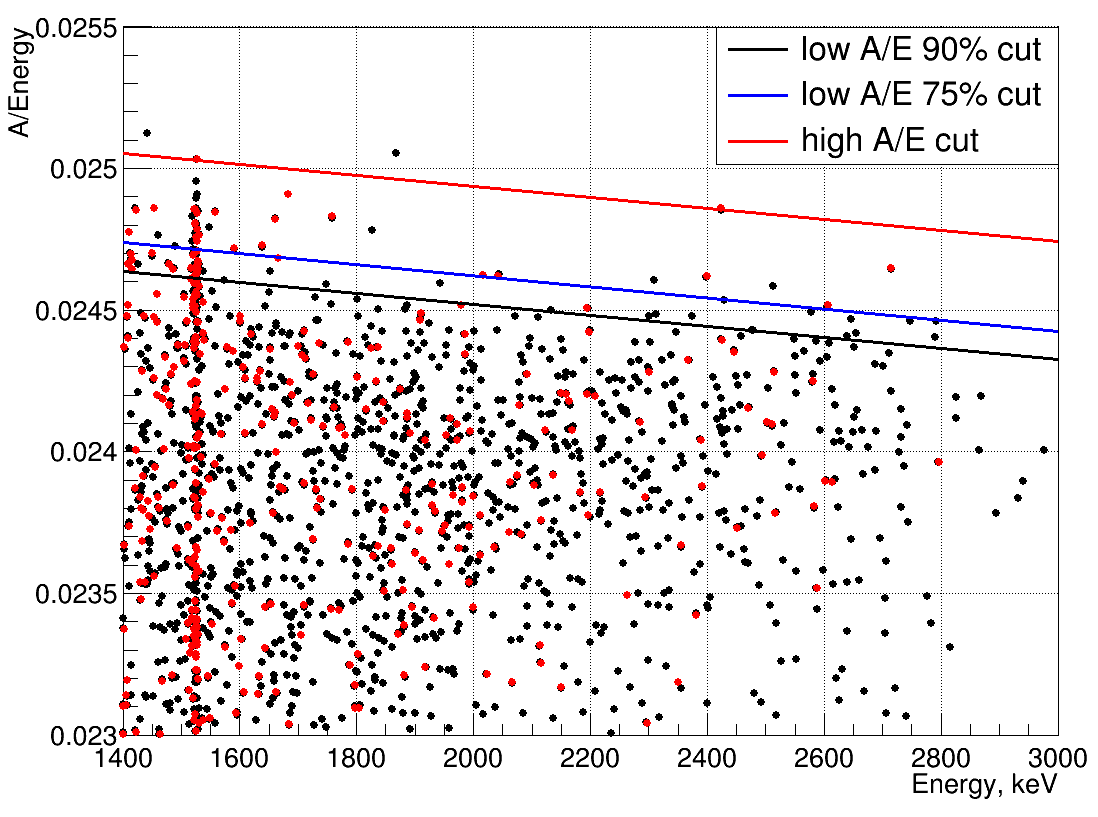}
     \caption{\label{fig:ae_pmt}A/E versus the event energy registered by the BEGe detector in \LArGe{} without NMS. Red dots are the events removed by coincidence with the LAr veto.}
\end{center}
\end{figure}

The first tests of \I{42}{K} suppression by PSD were performed with the bare BEGe detector in \LArGe{} without a mini-shroud with the smaller \I{42}{Ar} source. The A/E distribution is shown in Fig.~\ref{fig:ae_pmt}.
The majority of events coming from the beta decay of \I{42}{K} are located below the SSE band. They are predominantly NSP from electrons which deposit their energy in the n\ts{+} surface of the germanium detector. Only a small amount of the events are located in the SSE band. 

The LAr veto is not very efficient in suppressing events from \I{42}{K} in the ROI, because such electrons often do not deposit much energy in liquid argon. However, combined PSD and LAr veto work efficiently, helping to suppress different types of background. The LAr veto acceptance measured by pulse generator is 96.5\%. Part of the events in the SSE band are coming from other background sources (mostly from \I{228}{Th}). Such gamma background is removed with high probability by the scintillation veto~\cite{LArGe}. After applying PSD and the LAr veto the events in the ROI are removed almost completely.

Tab.~\ref{tab:suppress_bare} summarizes the SFs in different energy ranges.
\begin{table}
  \begin{center}
  \caption{\label{tab:suppress_bare}Number of counts from the \I{42}{K} measurements and suppression factors in two energy regions, before and after applying different cuts for the BEGe detector without the NMS.}
\begin{tabular}{lcc}
\toprule
 &\multicolumn{2}{c}{Energy region [keV]} \\
 & 1520 -- 1530 & 1839 -- 2239  \\
 &  & ROI \\
\hline
Events before cuts & 427 & 610  \\
~\textit{Including other backgr.} & \textit{0.9} & \textit{23} \\
After PSD + LAr veto & 10 & 2 \\
~\textit{Including other backgr.} & \textit{0.015} & \textit{0.62} \\
SF & 41.2 &  \(>\)121 \\
\bottomrule
\end{tabular}
  \end{center}
\end{table}
Only 2 events survive from the initial 610 events in the $\pm$200\,keV window around the ROI of \bbno{} after applying PSD and LAr veto. For a precise estimation of the \I{42}{K} SFs the remaining background sources have to be taken into consideration. Unfortunately, no measurements prior to dissolving of \I{42}{Ar} are available for this detector to estimate the background level. So it is estimated with help of previous investigations with an encapsulated semi-coaxial germanium detector. The background level of that detector was (0.12 -- 4.6)$\times$10$^{-2}$\,\ctsper{} after LAr veto~\cite{LArGe}. It was dominated by \I{208}{Tl} external to the cryostat, presumably from the PMTs~\cite{LArGe}, so we can expect to have this background with other detectors as well. In addition to this background the contribution of cosmogenic background from \I{68}{Ga} in the BEGe detector is taken into account. The background index after applying the LAr veto and PSD is estimated with help of previous investigations in \LArGe{}~\cite{LArGe} and simulations~\cite{phd_bjoern}. The accumulated statistics is not enough to find precisely the suppression factor for the \I{42}{K} background in the ROI, but it is possible to obtain a lower limit of $>$121 at 90\% C.I. for the combined LAr veto and PSD cuts. This value is obtained with help of TRolke class~\cite{TRolke} assuming poisson statistics of the signal. This value can be improved by applying a stronger PSD cut: after applying 73\% PSD cut (low 75\% and high cuts) no events survive the PSD veto (see Fig.~\ref{fig:ae_pmt}).

It is necessary to notice that the suppression factor of PSD depends on the detector performance: with worse A/E resolution some of the events can leak into the SSE band, reducing the efficiency of PSD cuts. That is why the SF achieved in \LArGe{} may be slightly different than for the detectors in \Gerda{}~Phase~II with different PSD performance.
%
\subsection{ \I{42}{K} background suppression of NMS combined with PSD and LAr veto}
\label{sec:invtot}
%

The performance of the LAr veto with introduced NMS was investigated with a \I{228}{Th} source and with spiked \I{42}{Ar} in liquid argon. The suppression factors of the LAr veto are obtained with the two versions of the NMSs and without it (see Tab.~\ref{tab:pmt_veto1}).
\begin{table}
  \begin{center}
    \caption{\label{tab:pmt_veto1}Suppression by the LAr veto for the measurements with and without \I{228}{Th} sources. The spiked \I{42}{Ar} was present in all measurements}
\begin{tabular}{lrrr} 
  \toprule
Energy & \multicolumn{3}{c}{Suppression factors} \\
regions & &  & \\
\ [keV] & without NMS & welded NMS &glued NMS \\
\hline
\multicolumn{4}{l}{Measurements without \(^{228}\)Th source} \\
1520 -- 1530 & 3.3(4) & 10.9(1.1) & 10.8(1.3)\\
1839 -- 2239 & 1.22(10) & 1.33(12) & 1.37(16) \\
\hline
\multicolumn{4}{l}{Measurements with external \(^{228}\)Th} \\
1839 -- 2239 & 26.8(19) & 37.2(19) & 31.9(8) \\
\bottomrule
\end{tabular}
  \end{center}
\end{table}
The LAr veto SFs improves when using the NMS, which is clearly seen in the suppression of the \I{42}{K} gamma line. The absorption length of light shifted by NMS is much longer compared to the 128\,nm scintillation light, and thus more light can reach the LAr veto light instrumentation. This may be a possible reason of the enhanced performance of the LAr veto.

\begin{figure}
\begin{center}
    \includegraphics[width=1\linewidth]{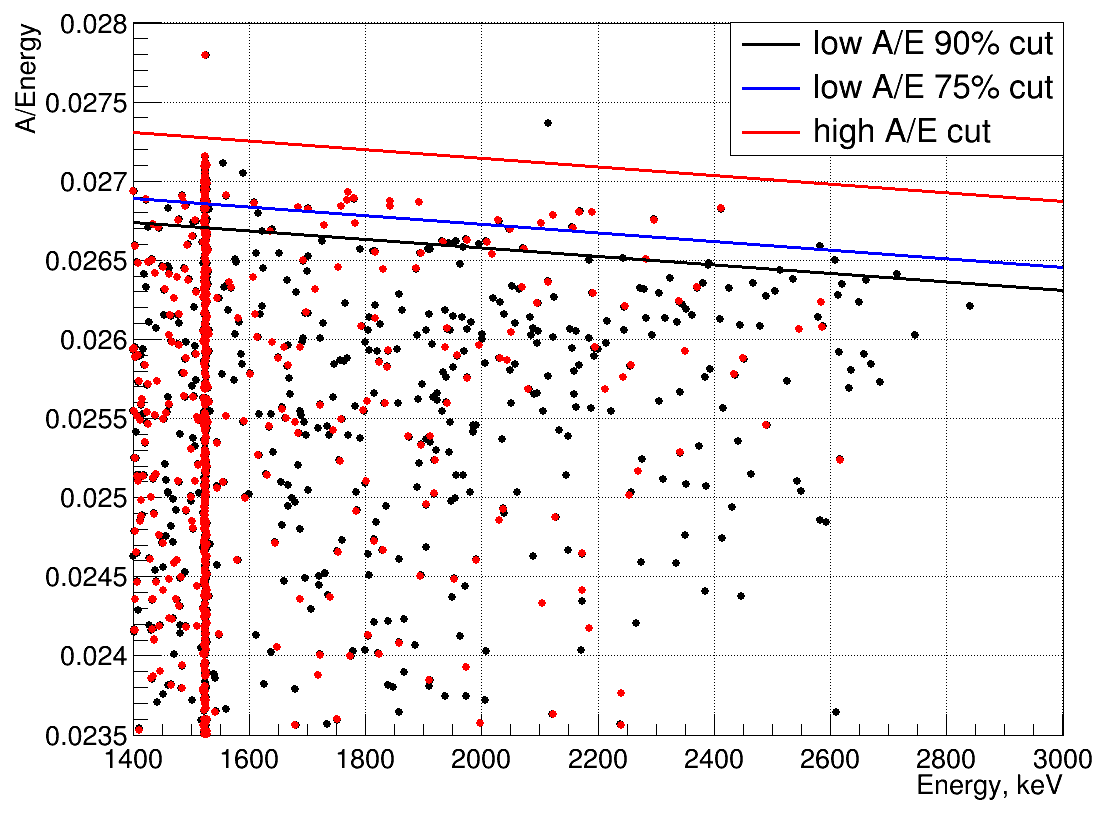}
     \caption{\label{fig:ae_pmt_nms}A/E versus the event energy registered by the BEGe detector in \LArGe{} with NMS. Red dots are the events removed by coincidence with the LAr veto.}
\end{center}
\end{figure}

The combined performance of the NMS, PSD and LAr veto is scrutinized by \I{42}{Ar} measurements of the BEGe detector with the NMS. The A/E distribution measured with the \I{42}{Ar} source and with LAr veto is shown in Fig.~\ref{fig:ae_pmt_nms}.
In this measurement the final, glued version of the NMS was implemented. Both \I{42}{Ar} sources were spiked into the \LArGe{} cryostat. Similarly to the previous measurements without NMS, most of the events coming from \I{42}{K} are rejected by PSD. The remaining part is removed by the LAr veto. The LAr veto allows to suppress events located in the SSE band more efficiently than in the MSE band (see Fig.~\ref{fig:ae_pmt_nms}). This is an indication that most of the events in the SSE band are not electrons from \I{42}{K}. The number of events surviving PSD and LAr veto are shown in Tab.~\ref{tab:suppress3}.
\begin{table}
  \begin{center}
    \caption{\label{tab:suppress3}Number of counts from the \I{42}{K} measurement, before and after applying LAr veto and PSD cuts for the BEGe detector with the NMS.}
\begin{tabular}{lcc}
\toprule
 & \multicolumn{2}{c}{Energy regions [keV]} \\
 & 1520 -- 1530 & 1839 -- 2239 \\[1mm]
 &  & ROI \\
\hline
Before cuts & 750 & 216 \\
~\textit{Including other backgr.} & \textit{0.6}  & \textit{20} \\
After PSD + LAr veto & 5 &2 \\
~\textit{Including other backgr.} & \textit{0.009} & \textit{0.4} \\
SF & 144 & \textgreater{38} \\
\bottomrule
\end{tabular}
  \end{center}
\end{table}
From the initial 216 events in the ROI only 2 events survive after applying LAr veto and PSD cuts. The estimated number of background events is 0.4. After taking into account the pulser acceptance, the SF is greater than 38 (90\% C.I.). By applying a stronger PSD cut (73\%) it is possible to suppress all of the events. 

\begin{figure}
  \begin{center}
    \includegraphics[width=1\linewidth]{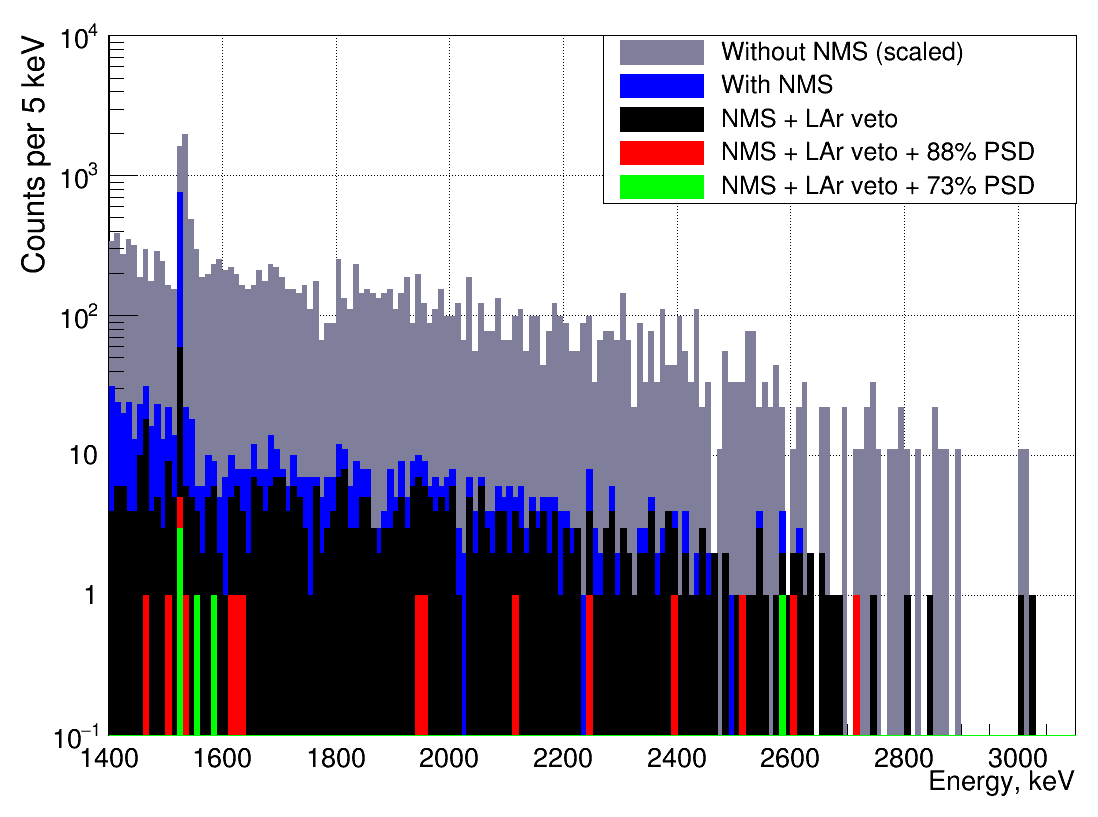}
    \caption{\label{fig:all_veto} Energy spectra demonstrate suppression of the events from \I{42}{Ar} by the NMS and the combined PSD and LAr cuts.}
  \end{center}
\end{figure}

Fig.~\ref{fig:all_veto} shows the comparison of spectra with and without NMS and the effect of active \I{42}{Ar} suppression. 
The gray area shows the spectrum of the bare BEGe detector without NMS scaled to the measurement time with the NMS. Other spectra demonstrate the \I{42}{K} suppression by the NMS and different cuts. In the scaled energy spectrum registered without the NMS there are 4384 events in ROI. After applying the PSD cut and the LAr veto only 2 events survive, including an estimated background of 0.4 events. Taking into account other than \I{42}{K} background components, LAr veto and the NMS suppression factor, the resulting suppression factor is higher than 513 at 90\% C.I. is obtained. (see Tab.~\ref{tab:bi}). 
\begin{table*}
  \begin{center}
  \caption{\label{tab:bi}Suppression factors in \LArGe{} and expected \I{42}{K} background rate in \Gerda{}~Phase~II like setup before and after applying the LAr veto and PSD cuts.}
\begin{tabular}{lrr}
\toprule
Experimental & Suppression & Expected \I{42}{K} background \\
conditions  & factor & in $\pm$200 keV ROI\\
  & in \LArGe{}  & [10$^{-3}$\,\ctsper] \\
\hline
no NMS  & 1& [210..800]\ts{1}\\
NMS &  14.3(2.1) & [13..66]\\
PSD + LAr veto (no NMS) & \textgreater{121} & \textless{[1.7..7]}\\
NMS + PSD + LAr veto &\textgreater{513} & \textless{[0.4..1.6]}\\  
NMS + PSD + LAr veto\ts{2} &\textgreater{1476} & \textless{[0.14..0.5]}\\  
\bottomrule
\end{tabular}
  \end{center}
{\footnotesize
\ts{1} Estimation is based on a simulation using a realistic range of dead layer thicknesses.\\
\ts{2} Estimation made with help of data taken without NMS, but with higher statistics}
\end{table*}

The measurement with NMS has lower counting statistics compared to measurements without NMS. However, there is no indication that PSD performance in case of NMS is worse than in the measurements without NMS. The LAr veto works slightly better in case of NMS, but it almost does not suppress events from \I{42}{K} in the ROI, that is why it has little effect on the results. Therefore it is safe to assume SF~\textgreater~121, obtained in the measurement without NMS and better statistics (see Tab.~\ref{tab:bi}) instead of SF~\textgreater~38. With the NMS suppression factor of 14.3(2.1) using SF~\textgreater~121 for PSD and LAr veto a total SF of greater than 1476 is obtained.  

%
\subsection{The NMS for \Gerda{} Phase II}
\label{sec:nmsGERDA}
\begin{figure}
  \begin{center}
    \center{\includegraphics[width=0.8\linewidth]{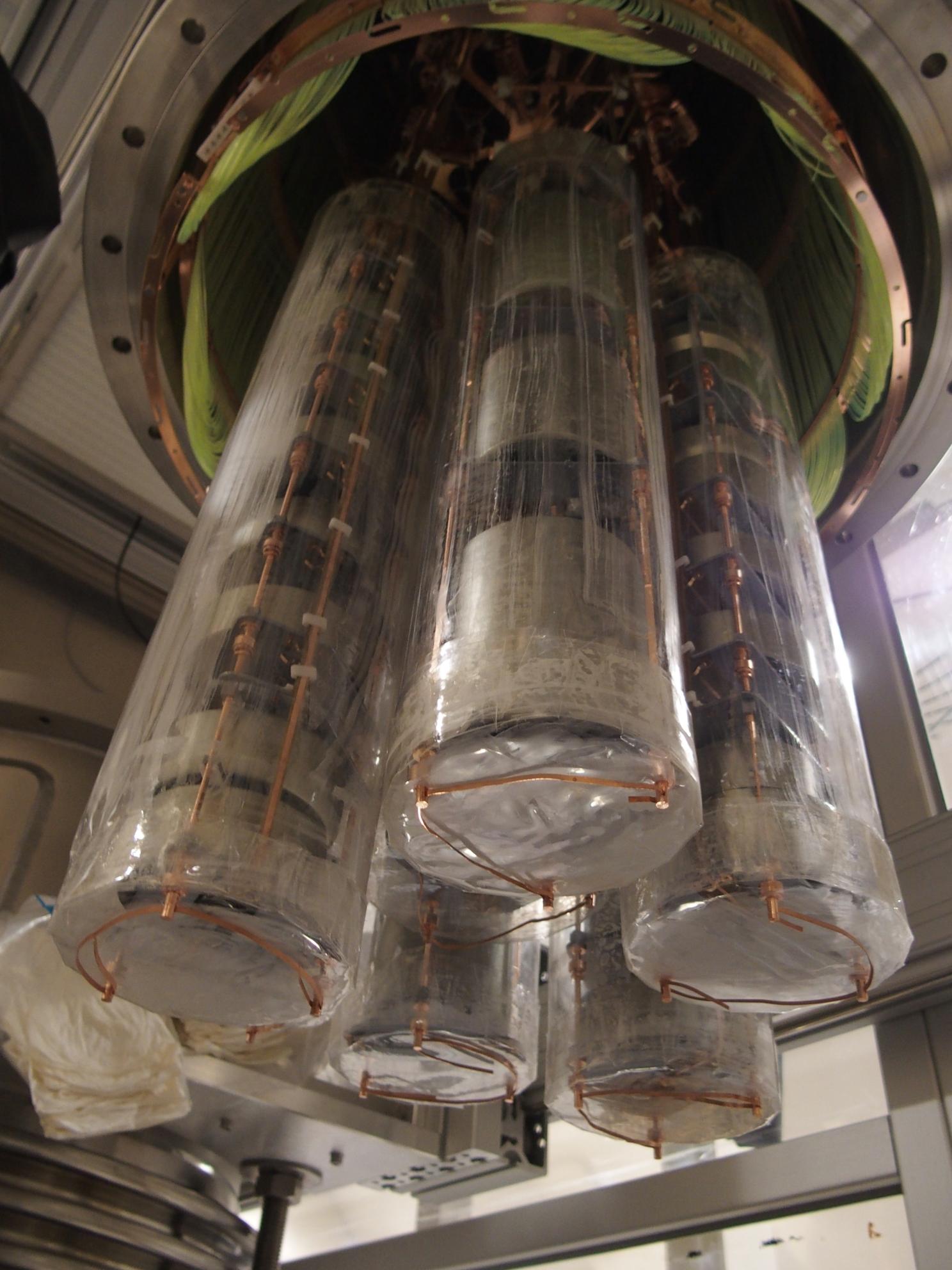}}
    \caption{\label{fig:nms_GERDA}Seven \Gerda{} detector strings with the NMSs prior to insertion into the cryostat}
  \end{center}
\end{figure}

%
For each detector string in \Gerda{} a dedicated NMS was prepared and mounted. A photo of the \Gerda{} Phase II integration is shown in Fig.~\ref{fig:nms_GERDA}. 
The setup was introduced into the \Gerda{} cryostat and data taking successfully started. The achieved background rate in \Gerda{}~Phase~II after all cuts is approximately 1$\times$10$^{-3}$ \ctsper, satisfying desired specifications~\cite{gerda_nature}. So the described concept of \I{42}{Ar} background suppression has proven its effectiveness. Using SFs obtained in \LArGe{} it is possible to estimate the \I{42}{K} background level in \Gerda{}~Phase~II like setup. 

The expected \I{42}{K} background rate in \Gerda{} is estimated by scaling the \I{42}{Ar} background in \LArGe{} to \Gerda{} using the ratio of the 1525 keV line intensities. The dead layer thickness has a big influence on the \I{42}{K} count rate in the ROI. That is why this factor is taken into consideration with the help of simulations. From these calculations it is expected that a bare BEGe detector in \Gerda{} without a mini-shroud and before applying any cuts has a \I{42}{K} BI in the range of [2.1..8]$\times$10$^{-1}$ \ctsper. The lower bound corresponds to a dead layer thickness of 1.0\,mm with minimal expectation for the \I{42}{K} concentration in liquid argon, the higher value corresponds to 0.6\,mm thickness and maximal expectation for the \I{42}{K} concentration. A typical value of the dead layer of a BEGe detector in \Gerda{} is about 0.8\,mm~\cite{GERDAbkg}, so the limit for the \I{42}{K} background rate is expected to lay in the quoted range. By applying the SF derived in \LArGe{} we obtain the background expectations in \Gerda{} like setup after NMS, PSD and LAr veto (see Tab.~\ref{tab:bi}).

However, it is necessary to note that obtained values just demonstrate possibility to suppress \I{42}{Ar} background to desired level, but can not be applied directly to \Gerda{} setup. \Gerda{} has 30 BEGe detectors and the LAr veto is rather different compared to \LArGe{}. In principle, the difference in the LAr veto should not affect much the total suppression of the \I{42}{K} background, while its suppression is low (factor $\sim$ 1.2) compared to NMS and PSD (see Tab.~\ref{tab:pmt_veto1}). But some of the detectors in \Gerda{} have a worse PSD performance than one tested in \LArGe, so a worse suppression by PSD is expected for these detectors. One the other hand the suppression factors of NMS in \Gerda{} should be higher due to the better design of the \Gerda{} NMS with less inner volume for each detector.

However, all these differences should not be very big and current calculation gives a good estimate about possible $^{42}$K background contribution for \Gerda{}~Phase~II.

\section{Conclusions}
\label{sec:conclusions}
%
Our investigation shows that the NMS alone is an efficient tool to suppress the background caused by \I{42}{K} about 15 times. It is also demonstrated that PSD is a very good method to mitigate surface events from \I{42}{K} decays. Results of the measurements in \LArGe{} with a combination of NMS show that the NMS together with the PSD and LAr veto can suppress \I{42}{K} background by more than 3 orders of magnitude.
  
The glued NMS is easy to construct and robust enough for handling and mounting. Moreover, there is a clear indication that NMS improves the performance of LAr veto. The NMS is made from radiopure material. The expected contribution to the background index from its contamination is only 4.6$\times$10$^{-4}$\,\ctsper{} and 5.3$\times$10$^{-6}$\,\ctsper{} before and after the LAr scintillation veto, respectively. These values are well below the requirements for \Gerda{}~Phase~II and may also be acceptable for future experiments.

The NMSs were successfully mounted and immersed into the liquid argon cryostat of \Gerda{}. The described concept of \I{42}{Ar} suppression proved its usability and performance. Taking into account the results obtained in \LArGe{} and simulations we estimate the possible level of background from \I{42}{K} in \Gerda{}~Phase~II. It is expected to be \textless[0.14..0.5]\(\times 10^{-3}\)\ctsper{} with NMS and after applying PSD and LAr veto cuts in the ROI. This limit already meets the Phase~II design criteria and can be further improved by a stronger PSD cut or by using the detectors with thicker dead layers.

\begin{acknowledgements}
We would like to thank F. Calaprice from the Princeton University for providing us the nylon. A special thank to Maria Laura di Vacri and Stefano Nisi from LNGS, for their involvement in the searches for radiopure components. Big thank to Bernhard Schwingenheuer for many suggestions and discussions. The investigations were supported by the Max Planck Society (MPG), the German Research Foundation (DFG) via the Excellence Cluster Universe, the Italian Istituto Nazionale di Fisica Nucleare (INFN), the Swiss National Science Foundation (SNF), the Russian Foundation for Basic Research – RFBR (grant 16-02-01096), the Polish National Science Centre (grant No. UMO-2012/05/E/ST2/02333) and the Foundation for Polish Science (grant No. TEAM/2016-2/17). 
\end{acknowledgements}



\end{document}